\documentclass[prl,superscriptaddress,twocolumn]{revtex4}
\usepackage[latin9]{inputenc}
\usepackage{color}
\usepackage{amsmath}
\usepackage{amssymb}
\usepackage{graphicx}
\usepackage[unicode=true,
 bookmarks=true,bookmarksnumbered=false,bookmarksopen=false,
 breaklinks=false,pdfborder={0 0 1},backref=false,colorlinks=true]
 {hyperref}
\hypersetup{
 linkcolor=magenta,urlcolor=blue,citecolor=blue,pdfstartview={FitH},urlcolor=blue}

\makeatletter
\usepackage{amsmath}
\usepackage{amssymb}
\usepackage{graphicx}
\usepackage{graphbox}

\usepackage{amsfonts}
\usepackage{tabularx}
\usepackage{dcolumn}
\usepackage{algorithm}  
\usepackage{algorithmicx}  
\usepackage{algpseudocode}  
\usepackage{bm}

\usepackage{epstopdf}
\usepackage{times}

\makeatother

\begin{document}
\title{Universal Adversarial Examples and Perturbations for Quantum Classifiers}
\author{Weiyuan Gong}
\affiliation{Center for Quantum Information, IIIS, Tsinghua University, Beijing
100084, People's Republic of China}

\author{Dong-Ling Deng}
\email{dldeng@tsinghua.edu.cn}
\affiliation{Center for Quantum Information, IIIS, Tsinghua University, Beijing
100084, People's Republic of China}
\affiliation{Shanghai Qi Zhi Institute, 41th Floor, AI Tower, No. 701 Yunjin Road, Xuhui District, Shanghai 200232, China}

\begin{abstract}
Quantum machine learning explores the interplay between machine learning and quantum physics, which  may lead to unprecedented perspectives for both fields. In fact, recent works have shown strong evidences that quantum computers could outperform classical computers in solving certain notable machine learning tasks. Yet,  quantum learning systems may also suffer from the vulnerability problem: adding a tiny carefully-crafted perturbation to the legitimate input data would cause the systems to make incorrect predictions at a notably high confidence level. In this paper, we study the universality of adversarial examples and perturbations for  quantum classifiers. Through concrete examples involving classifications of real-life images  and quantum phases of matter, we show that  there exist universal adversarial examples that can fool a set of different quantum classifiers. We prove that for a set of $k$ classifiers with each receiving input data of $n$ qubits, an $O(\frac{\ln k} {2^n})$ increase of the perturbation strength is enough to ensure a moderate universal adversarial risk. In addition, for a given quantum classifier we show that  there exist universal adversarial perturbations, which can be added to different legitimate samples and make them to be adversarial examples for the classifier.  Our results reveal the universality perspective of adversarial attacks for quantum machine learning systems, which would be crucial for practical applications of both near-term and future quantum technologies in solving machine learning problems. 
\end{abstract}

\maketitle

Machine learning, or more broadly artificial intelligence, has achieved dramatic success over the past decade \cite{Lecun2015Deep,Jordan2015Machine} and a number of problems that were notoriously challenging, such as playing the game of Go \cite{Silver2016Mastering, Silver2017Mastering} or predicting protein structures \cite{Senior2020Improved}, have been cracked recently. In parallel, the field of quantum computing \cite{Nielsen2010Quantum} has also made remarkable progress in recent years, with the experimental demonstration of quantum supremacy marked as the latest milestone \cite{Arute2019Quantum,Zhong2020Quantum}. The marriage of these two fast-growing fields gives birth to a new research frontier---quantum machine learning \cite{Biamonte2017Quantum,Dunjko2018Machine,Sarma2019Machine}. On the one hand, machine learning tools and techniques can be exploited to solve difficult problems in quantum science, such as quantum many-body problems \cite{Carleo2016Solving}, state tomography \cite{Torlai2018Neural},  topological quantum compiling \cite{Zhang2020Topological}, structural and electronic transitions in disordered materials \cite{Deringer2021Origins}, non-locality detection \cite{Deng2017MachineBN}, and classification of different phases of matter and phase transitions \cite{Zhang2017Quantum,Carrasquilla2017Machine,vanNieuwenburg2017Learning,Wang2016Discovering,
Broecker2017Machine,Chng2017Machine, Wetzel2017Unsupervised,Hu2017Discovering, Zhang2019Machine,Lian2019Machine}. On the other hand, new quantum algorithms running on quantum devices also possess the unparalleled potentials to enhance, speed up, or innovate machine learning \cite{Harrow2009Quantum,Lloyd2014Quantum,Dunjko2016QuantumEnhanced, Amin2018Quantum,Gao2018Quantum,Lloyd2018Quantum,Hu2019Quantum,Schuld2019Quantum}. Notable examples along this direction include the Harrow-Hassidim-Lloyd algorithm \cite{Harrow2009Quantum}, quantum principal component analysis \cite{Lloyd2014Quantum}, quantum generative models \cite{Gao2018Quantum,Lloyd2018Quantum,Hu2019Quantum}, and quantum support vector machines \cite{Rebentrost2014Quantum}, etc.  Without a doubt, the interaction between machine learning and quantum physics will benefit both fields \cite{Sarma2019Machine}.

\begin{figure}[h!]
\hspace*{-0.49\textwidth}
\includegraphics[width=.49\textwidth]{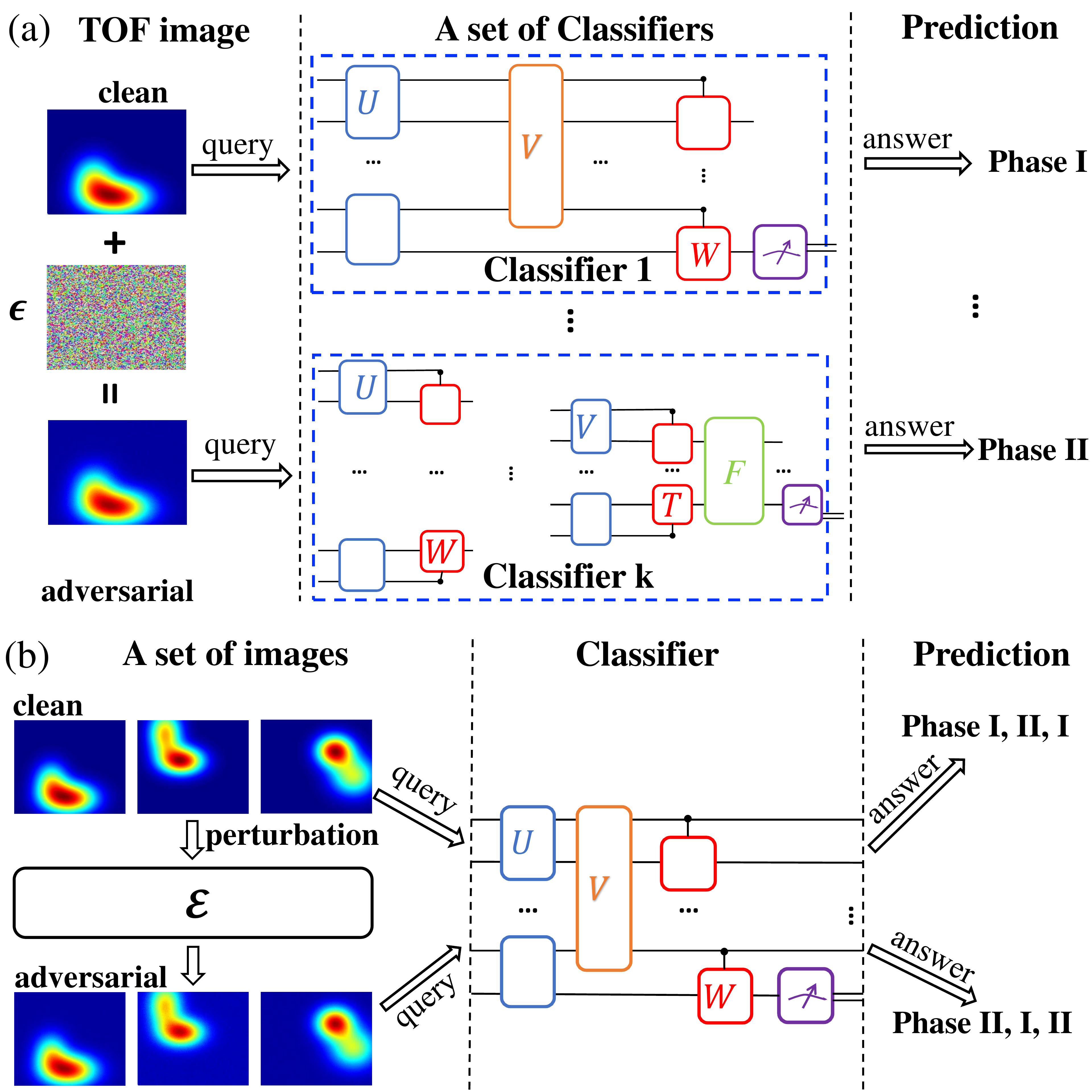}
\caption{A schematic illustration of universal adversarial examples and perturbations. (a) Universal adversarial examples: a set of quantum classifiers can be trained to assign phase labels to different time-of-flight images, which can be obtained directly in cold atom experiments. Adding a small amount of carefully crafted noise to a certain image could make it become a universal adversarial example, namely the new crafted image could deceive all the classifiers in the set.   (b)Universal adversarial perturbations: adding the same carefully-constructed noise to a set of images could make them all become adversarial examples for a given quantum classifier.}
\label{fig:main}
\end{figure}
In classical machine learning, it has been shown that classifiers based on deep neural networks are rather vulnerable in adversarial scenarios \cite{chakraborty2018adversarial,biggio2018wild,miller2019adversarial}: adding a tiny amount of carefully-crafted noises, which are even imperceptible to human eyes and ineffective to traditional methods, into the original legitimate data may cause the classifiers to make incorrect predictions at a notably high confidence level. A celebrated example that clearly showcases the vulnerability of deep learning was observed by Szegedy {\it et al.} \cite{Szegedy2014Intriguing}, where an image of a panda will be misclassified as a gibbon after adding an imperceptible amount of noises.  
The crafted input samples that would deceive the classifiers are called adversarial examples. Now, it is widely believed that the existence of adversarial examples is ubiquitous in classical machine learning---almost all learning models suffer from adversarial attacks, regardless of the input data types and the details of the neural networks \cite{chakraborty2018adversarial,biggio2018wild,miller2019adversarial}. More recently, the vulnerability of quantum classifiers has also been studied, sparking a new research frontier of quantum adversarial machine learning \cite{lu2020quantum,liu2019vulnerability,Du2020Quantum,Casares2020Quantum,Guan2020Robustness,Liao2020Adversarial}. In particular, Ref. \cite{lu2020quantum} explored different adversarial scenarios in the context of quantum machine learning and have demonstrated that, with a wide range of concrete examples, quantum classifiers are likewise highly vulnerable to crafted adversarial examples. This emergent research direction is growing rapidly, attracting more and more attentions across communities. Yet, it is still in its infancy and many important issues remain unexplored. 

In this paper, we consider such an issue concerning the universality of adversarial examples and perturbations for quantum classifiers. We ask two questions: (i) whether there exist universal adversarial examples that could fool a set of different quantum classifiers? (ii) whether there exist universal adversarial perturbations, which when added to different legitimate input samples could make them become adversarial examples for a given quantum classifier? Based on extensive numerical simulations and analytical analysis, we give affirmative answers to both questions. For (i), we prove that, by exploring the concentration of measure phenomenon \cite{ledoux2001concentration}, an $O(\frac{\ln k} {2^n})$ increase of the perturbation strength is enough to ensure a moderate universal adversarial risk for a set of $k$ quantum classifiers with each receiving input data of $n$ qubits; For (ii), we prove that, based on the quantum no free lunch theorem \cite{poland2020no,sharma2020reformulation}, the universal adversarial risk is bounded from both below and above and approaches unit exponentially fast as the number of qubits for the quantum classifier increase. We carry out extensive numerical simulations on concrete examples involving classifications of real-life images and quantum phases of matter to demonstrate how to obtain universal adversarial examples and perturbations in practice. 

\textit{Universal adversarial examples}.---To begin with, we first introduce some concepts and notations. 
Consider a classification task in the setting of supervised learning, where we assign a label $s\in S$ to an input data sample $\rho\in \mathcal{H}$, with $S$ being a countable label set and $\mathcal{H}$  the set of all possible samples. The training set is denoted as $\mathcal{S}_N=\{(\rho_1,s_1),...,(\rho_N,s_N)\}$, where $\rho_i\in \mathcal{H}$, $s_i\in S$, and $N$ is the size of the training set. 
Essentially, the task of classification is to learn a function (called a hypothesis function) $h:\mathcal{H}\rightarrow S$, which for a given input $\rho\in \mathcal{H}$ outputs a label $s$ \cite{goodfellow2016deep}. We denote the \textit{ground truth} function as $t:\mathcal{H}\rightarrow S$, which gives the true classification for any $\rho\in\mathcal{H}$. 
For the purpose in this paper, we suppose that after the training process the hypothesis function match the ground truth function on the training set, namely  $h(\rho)=t(\rho),\forall\rho\in \mathcal{S}_{N}$. We consider a set of $k$ quantum classifiers $\mathcal{C}_1,...,\mathcal{C}_k$ with corresponding hypothesis functions $h_i$ $(i=1,...,k)$ and introduce the following definitions to formalize our results.

\textit{Definition $1$.} We suppose the input sample $\rho$ is chosen from $\mathcal{H}$ according to a probability measure $\mu$ and $\mu(\mathcal{H})=1$. For $h_i$, we define $\mathcal{E}_i=\{\rho\in\mathcal{H}|h_i(\rho)\neq t(\rho)\}$ as the misclassified set, and the \textit{risk} for $\mathcal{C}_i$ is denoted as $\mu(\mathcal{E}_i)$.  

\textit{Definition $2$.} Consider a metric over $\mathcal{H}$ with the distance measure denoted as $D(\cdot)$. Then the $\epsilon$-expansion of a subset $\mathcal{H}'\subseteq\mathcal{H}$ is defined as: $
\mathcal{H}'_\epsilon=\{\rho|D_{\rm{min}}(\rho,\mathcal{H}')\leq\epsilon\}$, where $D_{\rm{min}}(\rho,\mathcal{H}')$ denotes the minimum distance between $\rho$ and any $\rho'\in\mathcal{H}'$. In the context of adversarial learning, a perturbation within distance $\epsilon$ added to the legitimate input sample $\rho\in\mathcal{E}_{i,\epsilon}=\{\rho'|D_{\rm{min}}(\rho',\mathcal{E}_i)\}$ can shift it to some misclassified one for the quantum classifier $\mathcal{C}_i$. Hence, we define the adversarial risk for $\mathcal{C}_i$ as $\mu(\mathcal{E}_{i,\epsilon})$. Similarly, the universal adversarial risk for a set of $k$ quantum classifiers is defined as $R=\mu(\mathcal{E}_\epsilon)$, where $\mathcal{E}_\epsilon=\cap_{i=1}^k\mathcal{E}_{i,\epsilon}$ denotes the set of universal adversarial samples. 

For technique simplicity and convenience, we focus on $\mathcal{H}=SU(d)$ (the special unitary group) with the Hilbert-Schmidt distance $D_{\text{HS}}(\rho,\rho')$ and Haar probability measure \cite{ratcliffe2006foundations}. We mention that the input data $\rho$ can be either classical or quantum in general. We treat both cases on the same footing since we can always encode the classical data into quantum states. We also note that any input state could be prepared by acting a unitary transformation on a certain initial state (e.g., the $|00\cdots 0\rangle$ state) and hence the classification of quantum states is in some sense equivalent to the classification of unitary transformations. Now, we are ready to present one of our main results.

\textbf{Theorem $1$.} Consider a set of $k$ quantum classifiers $\mathcal{C}_i$, $i=1,...,k$ and let $\mu(\mathcal{E})_{\text{min}}$ be the minimum risk among $\mu(\mathcal{E}_i)$. Suppose $\rho\in SU(d)$ and a perturbation $\rho\rightarrow \rho'$ occurs with $D_{\text{HS}}(\rho,\rho')\leq \epsilon$, then we can ensure that the universal adversarial risk  is bounded below by $R_0$  if 
\begin{equation}\label{eq:thm1}
\epsilon^2\geq\frac {4}{d}\ln{\left[\frac{2k}{\mu(\mathcal{E})_{\text{min}}(1-R_0)}\right]}.
\end{equation}

\begin{figure*}
\hspace*{-\textwidth}
\includegraphics[width=\textwidth]{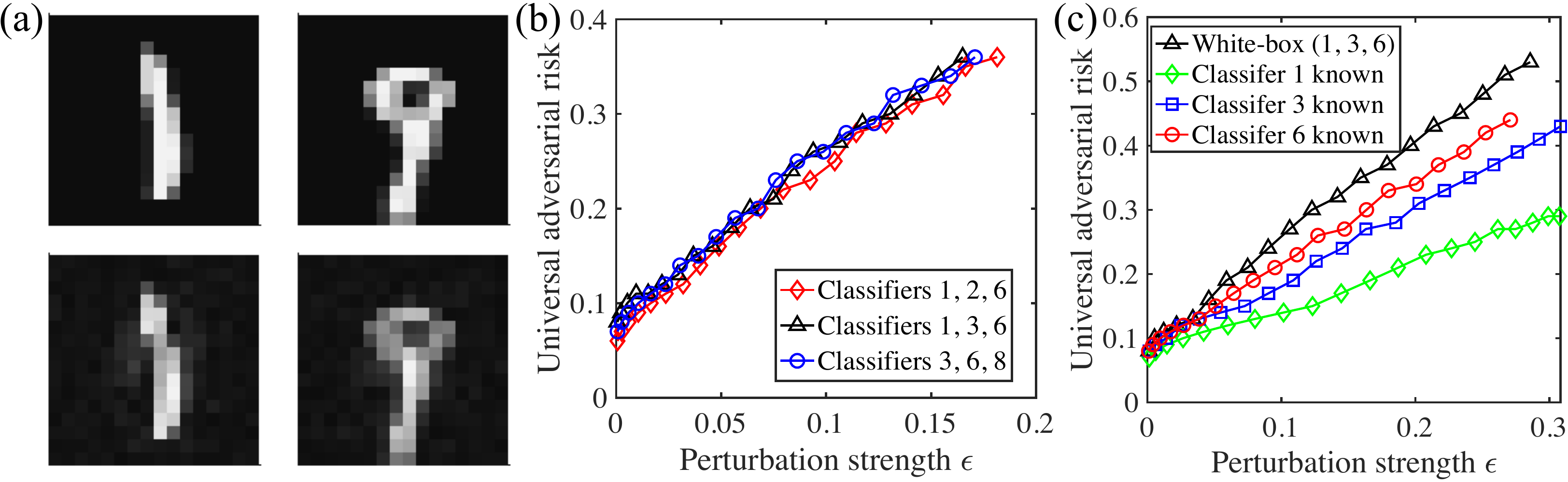}
\caption{Numerical results on universal adversarial examples. In this figure, the adversarial examples are obtained through the qBIM algorithm with step size $\alpha=0.02$.   (a) The clean and the corresponding universal adversarial handwritten digit images that can deceive all eight quantum classifiers. (b) The universal adversarial risk as a function of the perturbation strength $\epsilon$ for different subsets of the classifiers in classifying the ground states of the 1D transverse field Ising model. Here, we consider the white-box attack scenario and the universal adversarial risk is defined as the ratio of test samples that deceive all three classifiers in each subset. (c) Results for attacking a subset of classifiers consisting of classifiers $1$, $3$, and $6$, in a white-box black-box hybrid setting. Here, we assume that only one of the classifiers is known to the attacker, and for comparison the black curve with triangles plots the result for the white-box attack case. For more details, see the Supplementary Material  \cite{supplement}.}
\label{fig:Universal adversarial example}
\end{figure*}

\textit{Proof.} We give the main idea and intuition here. The full proof is a bit technically involved and thus left to the Supplementary Materials \cite{supplement}.  The first step is to prove that for a single quantum classifier $\mathcal{C}_i$, we can ensure that its adversarial risk is bounded below by $R_{0,i}$ if $\epsilon^2\geq\frac {4}{d}\ln{\left[\frac{2}{\mu(\mathcal{E}_i)(1-R_{0,i})}\right]}$. This can be done by exploring the concentration of the measure phenomenon for $SU(d)$ equipped with the Haar measure and Hilbert-Schmidt metric \cite{mahloujifar2019curse,liu2019vulnerability}.  Next, we use the De Morgan's laws in set theory to deduce that $\mu(\mathcal{E}_\epsilon)\geq 1-k+ \sum_{i=1}^k\mu(\mathcal{E}_i)$. In the last step, we choose $R_{0,i}=\frac{k-1+R_0}{k}$ and replace $\mu(\mathcal{E}_i)$ by $\mu(\mathcal{E})_{\text{min}}$ to increase $\epsilon$ a little bit for each $\mathcal{C}_i$. This leads to Eq. (\ref{eq:thm1}) and complete the proof. 

The above theorem implies that for a set of $k$ quantum classifiers with each receiving input data of $n$ qubits (thus $d=2^n$), an $O(\frac{\ln k}{2^n})$ increase of the perturbation strength would guarantee a moderate universal adversarial risk lower bounded by $R_0$. As $n$ increases, the lower bound of $\epsilon$ approaches zero exponentially. In other words, an exponentially small adversarial perturbation could result in universal adversarial examples that can deceive all $k$ classifiers with constant probability.  This is a fundamental feature 
of quantum classifiers in high dimensional Hilbert space due to the concentration of the measure phenomenon, independent of their specific structures and the input datasets.  

Although the above theorem indicates the existence of universal adversarial examples in theory, it is still unclear how to obtain these universal examples in practice. To deal with this issue, in the following we provide concrete examples involving classifications of hand-writing digit images and quantum phases with extensive numerical simulations. We mention that, in the classical adversarial machine learning literature, universal adversarial examples have also been shown to exist in real applications. For instance, in Ref. \cite{sharif2016accessorize} it is shown that an attacker can fool (such as dodging or impersonation) a number of the state-of-the-art face-recognition systems by simply wearing a pair of carefully-crafted eyeglasses. For our purpose, we consider a set of eight quantum classifiers with different structures, labeled by numbers from $1$ to $8$. The classifiers $1$ and $2$ are two quantum convolutional neural networks (QCNNs) \cite{cong2019quantum} and the classifiers $3-8$ are other typical multi-layer variational quantum circuits with depths from five through ten. The detailed descriptions of these quantum classifiers are given in the Supplementary Materials \cite{supplement}.

\begin{figure*}
\hspace*{-\textwidth}
\includegraphics[width=\textwidth]{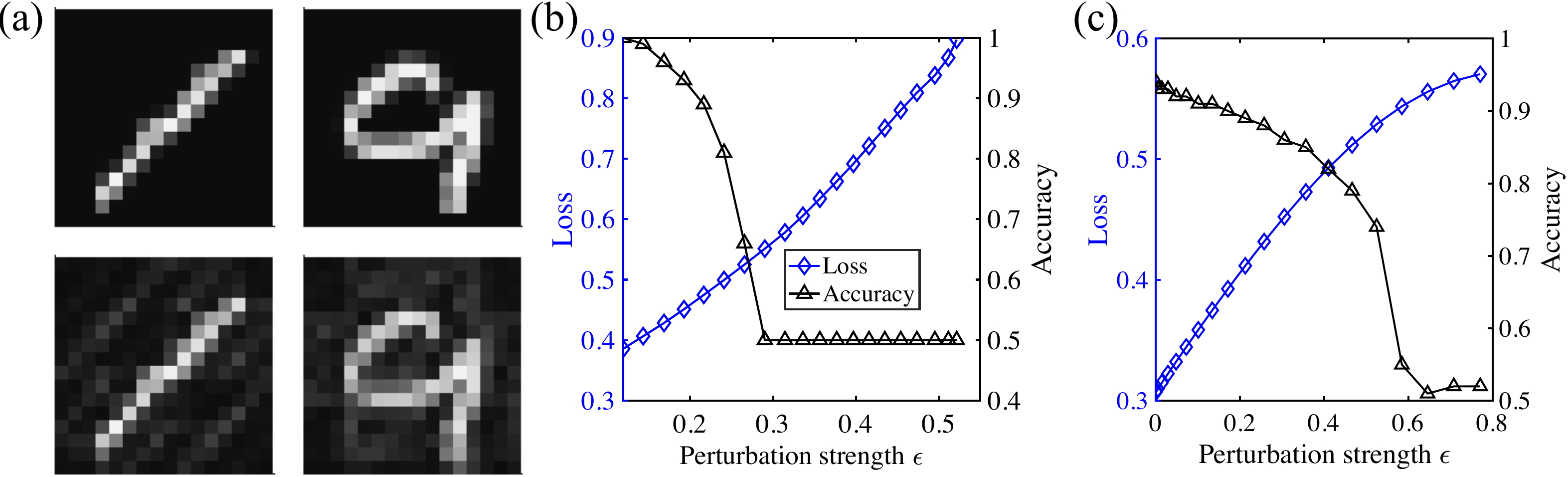}
\caption{Numerical results on universal adversarial perturbations. Similar to Fig. \ref{fig:Universal adversarial example}, in this figure the adversarial perturbations are also obtained by the qBIM algorithm with step size $\alpha=0.02$.  (a) The clean and  corresponding adversarial examples that can fool the quantum classifier $2$, which is a quantum convolutional neural network. These two adversarial images (bottom) are obtained by adding the same perturbation to the original legitimate ones.  (b) The loss and accuracy as  functions of the perturbation strength $\epsilon$ for the classifier $2$ in classifying the ground states of $H_{\text{Ising}}$. (c) A similar result for the classifier $8$. Throughout this figure, the white-box attack is considered. For more details, see the Supplementary Material  \cite{supplement}.}
\label{fig:exp2}
\end{figure*}

The first example we consider is the classification of handwritten-digit images in the MNIST dataset \cite{MNIST}, which is a prototypical testbed for benchmarking various machine learning scenarios. This dataset consists of gray-scale images of handwritten digits from $0$ through $9$, with each of them contains $28\times 28$ pixels. We reduce the size of the images to $16\times 16$, so that we can simulate the learning and attacking process of the quantum classifiers with moderate classical computational resources. We use amplitude encoding to map the input images into quantum states and the cross-entropy as the loss function for training and adversarial attacking. After training, we use the quantum-adapted basic iterative method (qBIM) \cite{Kurakin2016Adversarial}  to obtain the adversarial examples. The details of the training and adversarial attacking process are provided in the Supplementary Materials \cite{supplement}. In  Fig.\ref{fig:Universal adversarial example} (a), we display two universal adversarial examples for digits $1$ and $9$, which can deceive \textit{all} eight quantum classifiers at a high-confidence level. Notably, these universal adversarial examples only differ from the original legitimate ones slightly and they can be easily identified by human eyes. In fact,  the fidelity between the adversarial and legitimate samples is about $96\%$, which is fairly high given that the Hilbert dimension involved is  not very large ($d=256$ for this case). 

The above discussion concerns the vulnerability of quantum classifiers in classifying classical data (images). Yet, unlike classical classifiers that can only take classical data as input, quantum classifiers can  also directly classify quantum data (states) produced by quantum devices. To demonstrate the existence of universal adversarial examples in such a scenario, we consider classifying the ground state of the one-dimensional (1D) transverse field Ising model:
\begin{equation}\label{eq:Ising}
H_{\text{Ising}}=-\sum_{i=1}^{L-1}\sigma_i^z\sigma_{i+1}^z-J_x\sum_{i=1}^{L}\sigma_i^x,
\end{equation}
where $J_x$ denotes the strength of the transverse field and $\sigma_i^x$ and $\sigma_i^z$ are the Pauli matrices for the $i$-th spin. This Hamiltonian  maps to free fermions via Jordan-Wigner transformation \cite{sachdev2011quantum} and is exactly solvable. Its ground state features a quantum phase transition at $J_x=1$, between ferromagnetic phase with $J_x>1$ and paramagnetic phase with $0<J_x<1$. We consider classifying these two different phases by the eight quantum classifiers mentioned above, with the ground state as input data. We sample the Hamiltonian with varying $J_x$ from $0$ to $2$ and compute their corresponding ground states. These quantum states with their corresponding labels form the dataset required \cite{supplement}. 

In Fig. \ref{fig:Universal adversarial example}(b), we consider three subsets of quantum classifiers in a white-box attack setting (namely the attacker has full information about the learned model and the learning algorithm). We find that universal adversarial examples indeed exist for classifying quantum states, regardless of the internal structures of the classifiers.  As the perturbation strength $\epsilon$ increases, the universal adversarial risk increases roughly linearly with $\epsilon$. With a perturbation strength $\epsilon=0.18$, we find that $37\%$ of the test samples could become universal adversarial examples for each subset of the classifiers.  In Fig. \ref{fig:Universal adversarial example}(c), we consider a white-box black-box hybrid scenario, where the attacker knows only the full information  about  one classifier in the subset and does not have any information about the rest ones. The motivation of this consideration is to study the transferability of universal adversarial examples. From Fig. \ref{fig:Universal adversarial example}(c), we find that even with limited partial information, the adversary is still able to create universal adversarial examples, indicating a notable transferability property of these examples.  The universal adversarial risk also increases linearly with $\epsilon$, but it is noticeably smaller than that for the white-box case. This is consistent with the intuition that the more information the attacker has the easier to create adversarial examples. 

\textit{Universal adversarial perturbations}.---In the above discussion, we demonstrate, with both theoretical analysis and numerical simulations, that there exist universal adversarial examples that could deceive a set of distinct quantum classifiers. We now turn to the second question and show that there exist universal adversarial perturbations that can be added to different legitimate samples and make them adversarial to a given quantum classifier $\mathcal{C}$. Without loss of generality, we may consider a unitary perturbation $\hat{\epsilon}:\mathcal{H}\rightarrow\mathcal{H}$ as means of adversarial attack for all input samples. We denote the misclassified set as $\mathcal{E}=\{\rho\in\mathcal{H}|h(\rho)\neq t(\rho)\}$ and consequently the unitary adversarial set as $\mathcal{E}_{\hat{\epsilon}}=\{\hat{\epsilon}^{-1}(\rho)|\rho\in\mathcal{E}\}$. 

\textbf{Theorem $2$.} For an adversarial perturbation with unitary operator $\hat{\epsilon}$ and $n$ samples $\rho_1,...,\rho_n$ chosen from $\mathcal{H}$ according to the Haar measure, the performance of the quantum classifier $\mathcal{C}$ with $\hat{\epsilon}(\rho_1),...,\hat{\epsilon}(\rho_n)$ as input samples is bounded by: 
\begin{equation}
|R_E-\mu(\mathcal{E})|\leq \sqrt{\frac{1}{2n}\ln{(\frac2\delta)}}    \label{ReBound}
\end{equation}
with probability at least $1-\delta$ ($0<\delta<1$). Here $R_E$ is the empirical error rate defined as the ratio of the misclassified samples and $\mu(\mathcal{E})$ is the risk for $\mathcal{C}$. In addition, the expectation of the risk over all ground truth $t$ and training set $\mathcal{S}_N$ is bounded below by:
\begin{equation}\label{eq:thm2-2}
\mathbb{E}_t[\mathbb{E}_{\mathcal{S}_N}[\mu(\mathcal{E})]]\geq1-\frac{d'}{d(d+1)}(N^2+d+1), 
\end{equation}
where $d=\text{dim}(\mathcal{H})$ is the dimension of the input data and $d'=|S|$ is the number of output labels. 

\textit{Proof.}  We only sketch the major steps here and leave the details of the full proof to the Supplementary Materials \cite{supplement}. Noting that unitary transformations are invertible, the unitary perturbation operator $\hat{\epsilon}$ will transfer samples in $\mathcal{E}_{\hat{\epsilon}}$ into the misclassified set $\mathcal{E}$, and we can therefore deduce that $\mu(\mathcal{E})=\mu(\mathcal{E}_{\hat{\epsilon}})$. Then from the definition of $\mu(\mathcal{E})$, the Ineq. (\ref{ReBound}) follows straightforwardly by applying the Hoeffding's inequality \cite{hoeffding1963probability}. The derivation of the Ineq. (\ref{eq:thm2-2}) relies on the recent works about reformulation of the no free lunch theorem \cite{shalev2014understanding} in the context of quantum machine learning \cite{poland2020no,sharma2020reformulation} (see the Supplementary Materials for details).

This theorem indicates that in the limit $d\rightarrow \infty$, the expectation of the risk for a general quantum classifier goes to unit, independent of its structure and the training algorithm. For a fixed $d$, the lower bound of such an expectation decreases as  the number of the output labels or the size of the training set  increase. Adding an identical adversarial unitary perturbation to all possible data samples will not increase the risk on average. However, it is still possible for such a perturbation to increase the ratio of misclassified samples for a given finite set of $n$ original samples. In the following, we carry out numerical simulations and show how to obtain the universal adversarial perturbations in classifying images of handwritten digits and the ground states of the 1D transverse field Ising model. To implement the unitary perturbation $\hat{\epsilon}$, we add an additional variational layer  before the original quantum classifiers.  After training, we fix the variational parameters of the given classifier $\mathcal{C}$ and optimize the parameters of the perturbation layer through the qBIM algorithm to maximize the loss function for a given set of $n$ original samples \cite{supplement}.

The major results are shown in Fig. \ref{fig:exp2}. In Fig. \ref{fig:exp2}(a), we display two adversarial examples for digits $1$ and $9$, which are obtained by adding the same unitary perturbation to the original images and can fool the classifier $2$ (one of the QCNN classifiers mentioned above). We mention that the fidelity between the original and crafted images is relatively small (about $78\%$) compared with the examples given in Fig. \ref{fig:Universal adversarial example}(a), but the crafted images remain easily identifiable by human eyes. In Fig. \ref{fig:exp2}(b), we consider adding the same unitary perturbation to all the test samples of the ground states of $H_{\text{Ising}}$ in a white-box attack setting for classifier $2$. From this figure, it is clear that the accuracy drops rapidly at first as we increase the perturbation strength, and then maintains at a fixed finite value (about $0.5$). This is consistent with the Ineq.(\ref{ReBound}) that $R_E$ has an upper bound around $\mu(\mathcal{E})$. We mention that the loss keeps increasing as the perturbation strength increases, even in the region where the accuracy becomes flattened.  This counterintuitive behavior is due to the fact that the loss function (cross-entropy) is continuous, whereas the accuracy is defined by the ratio of correctly classified samples whose labels are assigned according to the largest output probability. Fig. \ref{fig:exp2}(c) shows similar results as in Fig. \ref{fig:exp2}(b), but for a different quantum  classifier (i.e., the classifier $10$ mentioned above).  

We remark that in our numerical simulations the Hilbert dimension involved is not very large due to limited classical computational resources. Consequently, a larger perturbation is needed to create the adversarial examples. As in Fig. \ref{fig:exp2}(a), the perturbation is perceptible to human eyes. However, this is by no means a pitfall in principle and can be circumvented by simulating larger quantum classifiers. As noisy intermediate-scale quantum devices \cite{Preskill2018Quantum} now become available in laboratories \cite{Arute2019Quantum}, this may also be resolved  by running the protocol in real quantum devices.  In addition, although we only focus on two-category classifications for simplicity in this paper, the extension to multi-category classifications and other adversarial scenarios is straightforward. 

\textit{Discussion and conclusion}.---This work only reveals the tip of the  iceberg in the fledgling field of quantum adversarial machine learning. Many important questions remain unexplored and demand further investigations. First, this work shows that the existence of universal adversarial examples is a fundamental feature of quantum learning in high-dimensional space in general. However, for a given learning task, the legitimate samples may only occupy a tiny subspace of the whole Hilbert space. This brings about the possibility of defending against adversarial attacks. In practice, how to develop  appropriate countermeasures feasible in experiments to strengthen the reliability of quantum classifiers still remains unclear. In addition,  unsupervised and reinforcement learning approaches may also suffer from the vulnerability problem \cite{vorobeychik2018adversarial}. Yet,  in practice it is often more challenging to obtain adversarial examples in these scenarios.  The study of quantum adversarial learning in the unsupervised or reinforcement setting is still lacking. In particular, how to obtain adversarial examples and perturbations and study their universality properties for quantum unsupervised or reinforcement learning remains entirely unexplored and is well worth future investigations.
Finally, it would be interesting and important to carry out an experiment to demonstrate the existence of universal adversarial examples and perturbations. This would be a crucial step toward practical applications of quantum technologies in artificial intelligence in the future, especially for  these applications in safety and security-critical environments, such as self-driving cars, malware detection, biometric authentication, and medical diagnostics \cite{finlayson2019adversarial}. 

In summary, we have studied the universality of adversarial examples and perturbations for quantum classifiers. We proved two relevant theorems: one states that an $O(\frac{\ln k}{2^n})$ increase of the perturbation strength is already sufficient to ensure a moderate universal adversarial risk for a set of $k$ quantum classifiers, and the other asserts that, for a general quantum classifier, the empirical error rate is bounded from both below and above and approaches to unit exponentially fast as the size of the classifier increases. We carried out extensive numerical simulations on concrete examples to demonstrate the existence of universal adversarial examples and perturbations for quantum classifiers in reality. Our results uncover a new aspect about the vulnerability of quantum machine learning systems, which would provide valuable guidance for practical applications of quantum classifiers based on both near-term and future quantum technologies. 

We thank Sirui Lu, Weikang Li, Xun Gao, Si Jiang, Wenjie Jiang and Nana Liu for helpful discussions. This work is supported by the start-up fund from Tsinghua University (Grant. No. 53330300320), the National Natural Science Foundation of China (Grant. No. 12075128), and the Shanghai Qi Zhi Institute.

\bibliographystyle{apsrev4-1-title}
\bibliography{WyGongbib}

\clearpage

\makeatletter
\setcounter{figure}{0}
\setcounter{equation}{0}
\renewcommand{\thefigure}{S\@arabic\c@figure}
\renewcommand \theequation{S\@arabic\c@equation}
\renewcommand \thetable{S\@arabic\c@table}

\begin{center} 
    {\large \bf Supplementary Material for: Universal Adversarial Examples and Perturbations for Quantum Classifiers}
\end{center} 

In this Supplementary Material, we provide more details about the proofs of the two theorems, structures of the quantum classifiers,  quantum encoding for classical data, training and attacking processes, and the algorithms for obtaining universal adversarial examples and perturbations.

\section{A. Proof for Theorem $1$}

In addition to the ones in the main text, we first give more notations and definitions to formulate the problem.

\textit{Definition A$1$.}  For $\mathcal{H}'\subseteq\mathcal{H}$, we define the \textit{concentration function} as $\alpha(\epsilon)=1-\inf\{\mu(\mathcal{H}'_\epsilon)|\mu(\mathcal{H})\geq\frac12\}$ with distance measure $D(\cdot)$ and probability measure $\mu(\cdot)$ in a $d$-dimensional vector space. 
If 
\begin{equation}\label{eq:def4}
\alpha(\epsilon)\leq \alpha e^{-\beta\epsilon^2d},
\end{equation}
then the vector space is said to be in $(\alpha,\beta)$-normal Levy group. 

We also introduce the following Lemma A1, which has already been obtained in  Ref. \cite{liu2019vulnerability}.  Here, we recap the statement and sketch the proof for completeness.

\textit{Lemma A$1$.} For a quantum classifier $\mathcal{C}_i$ that takes $\rho\in SU(d)$ according to the Haar measure $\mu(\cdot)$ as input and has a misclassified set $\mathcal{E}_i$. Suppose the adversarial input state $\rho'$ is restricted by $d_{HS}(\rho,\rho')\leq\epsilon$ to the clean data $\rho$. Then to guarantee an adversarial risk $R_i$, $\epsilon$ is bounded below by
\begin{equation}\label{eq:thm3}
\epsilon^2\geq\frac {4}{d}\ln{[\frac{2}{\mu(\mathcal{E}_i)(1-R_i)}]}.
\end{equation}

To prove Lemma A$1$, we further introduce the following two lemmas together with their brief proofs. 

\textit{Lemma A$2$. } (Theorem 3.7 in \cite{mahloujifar2019curse}) For each classifiers $\mathcal{C}_i$ and risk $\mu(\mathcal{E}_i)$, consider additional perturbation $\rho\rightarrow\rho'$, $\rho,\rho'\in \mathcal{H}$ and $D(\rho,\rho')\leq\epsilon$. 
If the adversarial risk $\mu(\mathcal{E}_{i,\epsilon})$ is guaranteed to be at least $R_i$, then $\epsilon^2$ must also be bounded by
\begin{equation}\label{eq:lem1}
\epsilon^2\geq\frac {1}{\beta d}\ln{[\frac{\alpha^2}{\mu(\mathcal{E}_i)(1-R_i)}]}.
\end{equation}

\textit{Proof. }We decompose the perturbation $\epsilon=\epsilon_1+\epsilon_2$. First construct a $\epsilon_1$ such that $\mu(\mathcal{E}_i)>\alpha e^{-\beta\epsilon_1^2d}$. Consider two cases for whether $\mu(\mathcal{E}_i)\leq\frac12$.

(i) If $\mu(\mathcal{E}_i)>\frac 12$, then we have $\mu(\mathcal{E}_{i,\epsilon_1})>\mu(\mathcal{E}_i)>\frac 12$.

(ii) If $\mu(\mathcal{E}_i\leq\frac12)$, suppose $\mu(\mathcal{E}_{i,\epsilon_1})\leq\frac12$. Then the complement probability $\mu(\mathcal{H}\backslash\mathcal{E}_{i,\epsilon_1})\geq\frac 12$. Denote $\mathcal{H}_i'=\mathcal{H}\backslash\mathcal{E}_{i,\epsilon_1}$, then $\mu(\mathcal{H}_i')\geq\frac12$ and $\mathcal{E}_i=\mathcal{H}\backslash\mathcal{H}_{i,\epsilon_1}'$. 
Hence, we can deduce a contradiction using \eqref{eq:def4} as $\alpha(\epsilon_1)\geq 1-\mu(\mathcal{H}'_{i,\epsilon_1})=\mu(\mathcal{E}_i)>\alpha(\epsilon_1)$.

Therefore, the perturbation $\epsilon_1$ ensures $\mu(\mathcal{E}_{i,\epsilon_1})>\frac 12$. Then we attach $\epsilon_2$ to $\mathcal{E}_{i,\epsilon_1}$, which is $\epsilon=\epsilon_1+\epsilon_2$ perturbation on $\mathcal{E}_i$. 
Applying \eqref{eq:def4} we can prove the lemma as $R_i=\mu(\mathcal{E}_{i,\epsilon})=\mu(\mathcal{E}_{i,\epsilon_1+\epsilon_2})>1-\alpha(\epsilon_2)$ and $\epsilon^2<\epsilon_1^2+\epsilon_2^2=\frac{1}{\beta d}\{\ln[\frac{\alpha}{\mu(\mathcal{E}_i)}]+\ln\frac{\alpha}{(1-R_i)}\}$.

\textit{Lemma A$3$. }$SU(d)$ group with Haar probability measure and normalized Hilbert-Schmidt metric is in $(\sqrt{2},\frac 14)$-normal Levy group \cite{gromov1983topological,giordano2007some}.

\textit{Proof.} First apply isoperimetric inequality \cite{gromov1983topological,milman2009asymptotic}, which states that for $\mathcal{H}'\subseteq\mathcal{H},\dim(\mathcal{H})=d$ and $\mu(\mathcal{H}')\geq\frac 12$,
\begin{equation}\label{eq:lem2-1}
\mu(\mathcal{H}'_\epsilon)\geq 1-\sqrt{2}e^{-\epsilon^2dR(\mathcal{H})/[2(d-1)]},
\end{equation}
where $R(\mathcal{H})=\inf_v{\text{Ric}(v,v)}$ for the Ricci curvature $\text{Ric}(v,v')$ of $\mathcal{H}$ and $v$ goes through all unit tangent vectors in $\mathcal{H}$. 
Combining \eqref{eq:lem2-1} and \eqref{eq:def4} we can deduce that
\begin{equation}\label{eq:lem2-2}
\alpha(\epsilon)\leq\sqrt{2}e^{-\epsilon^2dR(\mathcal{H})/2(d-1)}.
\end{equation}

According to \cite{meckes2014concentration}, for $SU(d)$ equipped with Hilbert-Schmidt metric, ${\rm Ric}(v,v)=\frac d2 G(v,v)$. And $G(v,v)$ is the Hilbert-Schmidt metric and $v$ is any unit tangent vector in $SU(d)$. Then from \cite{oszmaniec2016random} $G(v,v)=1$. Therefore, $R(\mathcal{H})=\frac d2$. This indicates that we can rewrite \eqref{eq:lem2-2} as
\begin{equation}\label{eq:lem2-3}
\alpha(\epsilon)\leq\sqrt{2}e^{-\epsilon^2d^2/4(d-1)}<\sqrt{2}e^{-\epsilon^2d/4}.
\end{equation}

Combining \eqref{eq:lem1} and \eqref{eq:lem2-3}, it is shown that for a classifier $\mathcal{C}_i$ with risk $\mathcal{E}_i$ which takes $\rho\in SU(d)$ as input and the Hilbert-Schmidt metric, to bound above adversarial risk with $R_i$, the adversarial perturbation is bounded below by $\epsilon^2\geq\frac {4}{d}\ln{[\frac{2}{\mu(\mathcal{E}_i)(1-R_i)}]}$. Hence, we have completed the proof for Lemma A1.

Now, we continue to prove the Theorem $1$ in the main text by using the Ineq. \eqref{eq:thm3}. We consider a set of quantum  classifiers  $\mathcal{C}_i,i=1,...,k$ with risk $\mathcal{E}_i,i=1,...,k$. Our goal is to calculate $\mu(\mathcal{E}_\epsilon)$ for a given $\epsilon$ perturbation. Consider the set $\mathcal{E}_{\text{set}}=\cap_{i=1}^k\mathcal{E}_i$ of original data that is misclassified by all classifiers in the set. If we assume an additional condition $\mathcal{E}_{\text{set}} \neq \emptyset$, 
then we can construct a quantum classifier $\mathcal{C}^*$ that misclassifies all $\rho\in\mathcal{E}_{\text{set}}$ and correctly classifies other states in $\mathcal{H}$. Then we apply \eqref{eq:thm3} to this classifier $\mathcal{C}^*$ and can deduce that to guarantee a  risk larger than $R_0$, the perturbation is bounded below by
\begin{eqnarray}
\epsilon^2\geq\frac {4}{d}\ln{[\frac{2}{\mu(\mathcal{E}_\text{set})(1-R_0)}]}. \label{SingleClass}
\end{eqnarray}

If the additional constraint is not satisfied, i.e. $\cap_{i=1}^k\mathcal{E}_i=\emptyset$, then we can not directly construct a quantum classifier $\mathcal{C}^*$. In this case, we notice that $\mathcal{E}_\epsilon=\cap_{i=1}^k\mathcal{E}_{i,\epsilon}=\mathcal{H}-\cup_{i=1}^k(\mathcal{H}-\mathcal{E}_{i,\epsilon})$. Therefore $\mu(\mathcal{E}_\epsilon)$ can be bounded below by:
\begin{equation}\label{eq:IEsize}
\mu(\mathcal{E}_\epsilon)\geq 1-\sum_{i=1}^k\frac{|\mathcal{H}\backslash\mathcal{E}_{i,\epsilon}|}{|\mathcal{H}|}=\sum_{i=1}^k\mu(\mathcal{E}_{i,\epsilon})-(k-1).
\end{equation}

Hence, if we attach a perturbation that ensures  $\mu(\mathcal{E}_{i,\epsilon})\geq R_{0,i}=\frac{k-1+R}{k}$ for each classifier $\mathcal{C}_i$, then the universal adversarial risk will be bounded below by $R$. Replacing $R$ and $\mu(\mathcal{E}_i)$ in \eqref{eq:thm3} with $R_0$ and $\mu(\mathcal{E})_{\text{min}}$, we finish the proof by arriving at the inequality:
\begin{equation}
\epsilon^2\geq\frac {4}{d}\ln{[\frac{2k}{\mu(\mathcal{E})_{\text{min}}(1-R_0)}]}. \label{Theorem1Supp}
\end{equation}

It is  worthwhile to mention that the Ineq. (\ref{Theorem1Supp}) holds regardless of whether the additional assumption $\mathcal{E}_{\text{set}} \neq \emptyset$ is satisfied or not. When $\mathcal{E}_{\text{set}} \neq \emptyset$ is satisfied, the problem reduces to the case for the single classifier $\mathcal{C}^*$. Yet, we cannot tell which inequality, either Ineq. (\ref{SingleClass}) or (\ref{Theorem1Supp}),  gives a tighter bound as we have no information about the value of $\mu(\mathcal{E}_{\text{set}})$ and $\mu(\mathcal{E}_{\text{min}})$. 
In our numerical simulations, among the test set containing 100 ground states of the Ising model, we find that there are five samples that can be misclassified by all eight quantum classifiers without adding any perturbation. This indicates that the additional condition might be satisfied in practice.

\section{B. Proof for Theorem $2$}

In this section, we provide the details of the proof for Theorem $2$ with some further discussions. Following the definitions in the main text, the adversarial operator $\hat{\epsilon}$ is unitary, and hence $\hat{\epsilon}^{-1}$ is also unitary. 
Then by applying the property of unitary transformation, we have
\begin{equation}\label{eq:eqsize}
\mu(\mathcal{E}_{\hat{\epsilon}})=\frac{|\hat{\epsilon}^{-1}(\mathcal{E})|}{|\mathcal{H}|}=\frac{|\mathcal{E}|}{|\mathcal{H}|}=\mu(\mathcal{E}).
\end{equation}
This indicates that the adversarial risk remains the same after we perform the same unitary perturbation operation $\hat{\epsilon}$ on every input quantum state $\rho\in\mathcal{H}$. 
 
We randomly pick $\rho\in\mathcal{H}$ according to the Haar measure. For each selection, the probability of misclassification occurrence is $\mu({\mathcal{E}_{\hat{\epsilon}}})=\mu(\mathcal{E})$. Therefore, we can regard each selection as a random variable, which will be $1$ when misclassification occurs and $0$ otherwise. 
Then, we apply Hoeffding's inequality for independent Bernoulli random variables and get with probability at least $1-\delta$ ($\delta>0$)
\begin{equation}
|R_E-\mu(\mathcal{E})|\leq\sqrt{\frac{1}{2n}\ln{(\frac2\delta)}}. 
\end{equation}
This proves the first part of the Theorem $2$ in the main text. 

To obtain an lower bound for $\mu(\mathcal{E})$, we further resort to the no free lunch theorem \cite{shalev2014understanding} and its reformulation in the context of quantum machine learning \cite{poland2020no,sharma2020reformulation}. Unlike in Ref. \cite{poland2020no}, where quantum input and  output are considered, our discussion is restricted to classification problems in which the output is classical labels. To this end,  here we give a loose estimation for the lower bound of $\mu(\mathcal{E})$ with some additional constraints according to our numerical simulations.

In our consideration, the quantum classifiers takes two steps to classify input samples. In the first step, the classifier takes a quantum state $\rho\in\mathcal{H}$ as input and undergoes a variational circuit to arrive at the output state $\rho_{\text{out}}$  belonging to a $d'$-dimensional Hilbert space. In the second step, the classifier outputs a label $s\in\{0,1,...,d'-1\}$ according to the largest probability among $\langle 0|\rho_{\text{out}}|0\rangle,\langle 1|\rho_{\text{out}}|1\rangle,...,\langle d'-1|\rho_{\text{out}}|d'-1\rangle$. Based on this, our analysis of $\mu(\mathcal{E})$ will lead to an average performance bound for the classifier \cite{poland2020no}.

In the first step from  $\rho$ to $\rho_{\text{out}}$, the  quantum ground truth is defined as a unitary process $t$. Without loss of generality, we may restrict our discussion to  the case of quantum pure states. The training set is rewritten as $\mathcal{S}_N=\{(|\psi_1\rangle,|\phi_1\rangle),...,(|\psi_N\rangle,|\phi_N\rangle)\}$ and the classifier learns a hypothesis operator $V$, which is a unitary process such that $t|\psi_i\rangle=V|\psi_i\rangle=|\phi_i\rangle$ for the training set. The  quantum risk function  is  defined as \cite{monras2017inductive}.
\begin{equation}\label{eq:qnfl-risk}
R_t(V)\equiv\int d|\psi\rangle||t|\psi\rangle\langle\psi|t^{\dagger}-V|\psi\rangle\langle\psi|V^{\dagger}||_1^2,
\end{equation}
where $||A||_1$ is the trace norm  for matrices \cite{Nielsen2010Quantum}. Now the quantum no free lunch theorem is described as below.

\textit{Lemma B$1$.(Quantum No Free Lunch)} The quantum risk function in a classification task averaged over selection of quantum ground truth $t$ and training set $S_N$ with respect to the Haar measure can be bounded below by
\begin{equation}\label{eq:thm4}
\mathbb{E}_t[\mathbb{E}_{\mathcal{S}_N}[R_t(V)]]\geq1-\frac{1}{d(d+1)}(N^2+d+1). 
\end{equation}

The  proof  of this lemma  and more discussions about its implications are provided in  Refs. \cite{poland2020no,sharma2020reformulation}. 
Here, we use this lemma to obtain Ineq. (4) in the main text. 
Noting that $||A||_1\leq 1$, hence for all the $\rho=|\psi\rangle\langle\psi|\in \mathcal{E}$, $D(t|\psi\rangle,V|\psi\rangle)=||t|\psi\rangle\langle\psi|t^{\dagger}-V|\psi\rangle\langle\psi|V^{\dagger}||_1\leq 1$. This means that $R_t(V)\leq 1$, regardless of whether the quantum data is correctly classified or not.

Then we come to the case when a quantum input is classified correctly. Without loss of generality, we can assume that the ground truth gives true label and output state $t|\psi\rangle=|i\rangle$, then since the quantum data is correctly predicted, $\langle i|t|\psi\rangle\langle\psi|t^{\dagger}|i\rangle\geq\frac{1}{d'}$. From this inequality, we obtain that the fidelity $F(V|\psi\rangle,t|\psi\rangle=|i\rangle)\geq\sqrt{\frac{1}{d'}}$. We can utilize the relation between fidelity and the trace norm
\begin{equation}\label{eq:two-distance}
D(\rho,\sigma)^2\leq 1-F(\rho,\sigma)^2,
\end{equation}
where $\rho,\sigma$ denote arbitrary quantum states. 

Hence,  for correctly classified quantum data we have $R_t(V)=D(t|\psi\rangle,V|\psi\rangle)^2=||t|\psi\rangle\langle\psi|t^{\dagger}-V|\psi\rangle\langle\psi|V^{\dagger}||_1^2\leq 1-F(t|\psi\rangle,V|\psi\rangle)^2\leq 1-\frac {1}{d'}$. As a result, the integral in Eq. \eqref{eq:qnfl-risk} is bounded by
\begin{equation}\label{eq:risk-rela}
R_t(V)\leq\mu(\mathcal{E})+\frac{d'-1}{d'}(1-\mu(\mathcal{E}))=\frac{1}{d'}(d'-1+\mu(\mathcal{E})).
\end{equation}

Combining \eqref{eq:thm4} and \eqref{eq:risk-rela}, we obtain a lower bound of $\mu({\mathcal{E}})$ averaged over ground truth $t$ and training set $\mathcal{S}_N$
\begin{equation}
\mathbb{E}_t[\mathbb{E}_{\mathcal{S}_N}[\mu(\mathcal{E})]]\geq1-\frac{d'}{d(d+1)}(N^2+d+1).
\end{equation}

This gives the Ineq. \eqref{eq:thm2-2} and complete the proof of Theorem $2$.

\begin{figure*}
\hspace*{-0.48\textwidth}
\includegraphics[width=.48\textwidth]{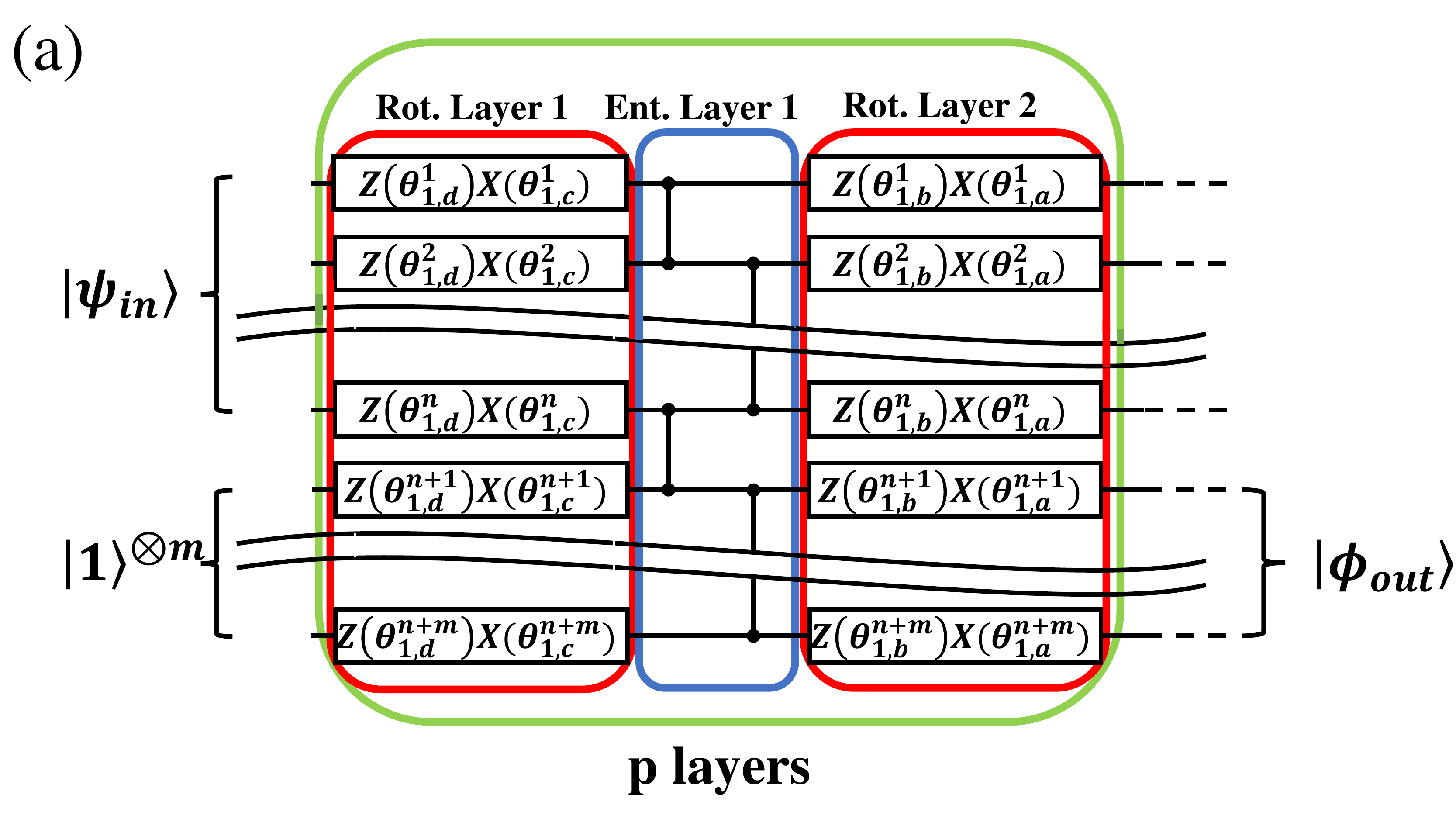}
\includegraphics[width=.48\textwidth]{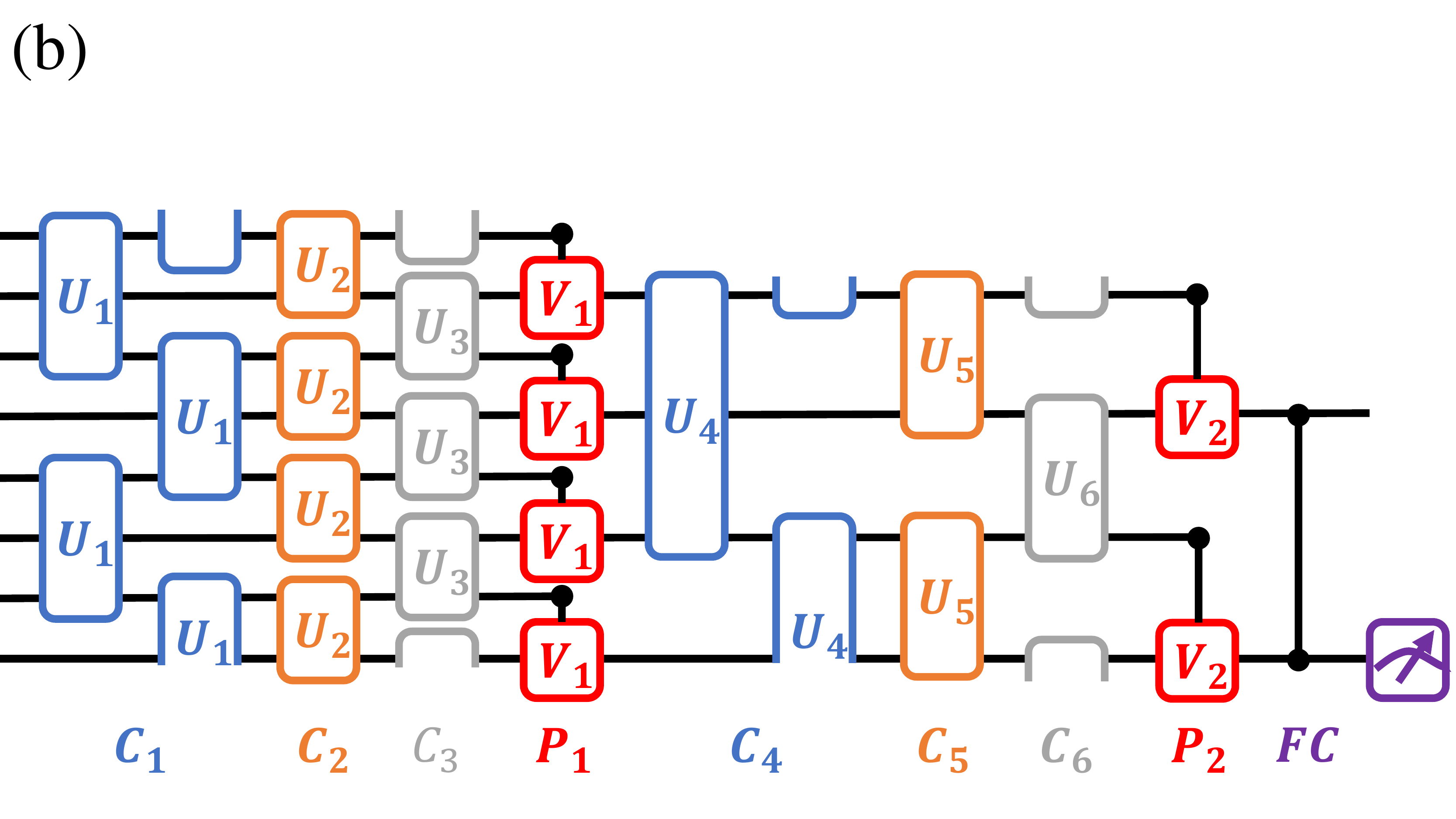}
\caption{The structure of quantum classifiers used in the numerical simulations.
(a)The illustrative structure of a general multi-layer quantum variational classifier that takes $n$-qubit  state $|\psi_{\text{in}}\rangle$ as input and outputs a $m$-qubit state $|\phi_{\text{out}}\rangle$. The classifier consists of $p$ layers and each layer consists two rotation units and an entangler unit. 
Each rotation unit contains a Euler rotation  $Z(\theta_{i,u}^{k})X(\theta_{i,v}^{k})$ [$(u,v)=(d,c)$ or $(b,a)$], where $i=1,...,p$ refer  to the number of layers, $k=1,...,m+n$ denote the number of qubit. After obtaining the output state $|\phi_{\text{out}}\rangle$, we compute the probabilities of projection measurements to predict and assign a label that corresponds to the largest probability.
(b) The illustrative structure of the QCNN classifier. This circuit contains six convolutional layers labeled by $C_1$ to $C_6$, two polling layers labeled by $P_1$ and $P_2$ respectively,  and a fully connected layer labeled by $FC$. The initial parameters are set to random values at the beginning of the training process. }
\label{qcircuit}
\end{figure*}

\section{C. The structures of quantum classifiers  and Encoding Methods}

\subsection{I. The structures of quantum classifiers}

In recent years, a number of  different quantum classifiers have been proposed \cite{schuld2020circuit,farhi2018classification,schuld2017implementing,mitarai2018quantum,li2017hybrid,Schuld2019Quantum,havlivcek2019supervised,zhu2019training,cong2019quantum,wan2017quantum,grant2018hierarchical,du2018implementable,uvarov2020machine,Rebentrost2014Quantum,blank2020quantum,tacchino2019artificial}. 
Here, we choose some of these classifiers to form the classifier set considered in this paper. As mentioned in the main text, our classifier set contains two QCNNs \cite{cong2019quantum} and six general multi-layer variational classifiers\cite{schuld2020circuit,farhi2018classification,li2017hybrid,mitarai2018quantum}. The sketch of a quantum variational circuit is shown in Fig. \ref{qcircuit}(a). 

In such a variational circuit model, we first prepare the $m+n$ qubit input state to be $|\psi_{\text{in}}\rangle\otimes|1\rangle^{\otimes m}$, where $|\psi\rangle_{\text{in}}$ is an $n$-qubit state that encodes the complete information of input sample to be classified. Then we apply a unitary transformation, which is composed of $p$ layers of interleaved operations, on the state. In each of the $p$ layers, there are two rotation units each performs arbitrary Euler rotations in Bloch sphere and an entangler unit consisting of  CNOT gates between each pair of neighboring qubits. The adjustable parameters are the rotation angles and are collectively denoted as $\Theta$. This generates a variational state:
\begin{equation}
|\Phi(\Theta)\rangle=\prod_{i=1}^pU_i(|\psi_{\text{in}}\rangle\otimes|1\rangle^{\otimes m}),
\end{equation}
where $U_i=\prod_k Z(\theta^k_{i,d})X(\theta^k_{i,c})U_{\text{ent}}Z(\theta^k_{i,b})X(\theta^k_{i,a})$ denotes the unitary operation for the $i$-th layer, with $U_{\text{ent}}$ representing the unitary operation generated by the entangler unit.

The brief structure of the QCNN and its hyperparameters utilized in this paper is shown in Fig. \ref{qcircuit}(b). The structure of the QCNN is the same as in Ref. \cite{cong2019quantum}.  

In our numerical simulations, we only focus on two-category classification problems. Thus, we only need one qubit to encode the labels $y=0,1$. After the variational circuits, the state of the output qubits becomes $\rho_{\text{out}}$. We compute $\mathbb{P}(y=m)=\text{Tr}(\rho_{\text{out}}|m\rangle\langle m|)$ and then assign  $y=1$ if $P(y=1)\geq P(y=0)$ and $y=0$ for other cases.

\subsection{II. Quantum encoding for classical data}

In the main text, one of the numerical simulations we did is based on the images of handwritten digits. In this dataset, the images are encoded classically, i.e. the data is encoded into a $m$-dimensional vector $\mathbf{v}$ in $\mathbb{R}^m$. To make such classical data processable to quantum classifiers, we need to convert the classic vector into a $n$-qubit quantum (pure) state in a $d=2^n$ dimensional Hilbert space. This converting process is called a quantum encoder. In this paper, we use the amplitude encoder to transfer classical data into quantum states \cite{schuld2020circuit,schuld2017implementing,Harrow2009Quantum,cong2016quantum,Rebentrost2014Quantum,kerenidis2017quantum,giovannetti2008architectures,lloyd2013quantum,wiebe2014quantum,giovannetti2008quantum,Aaronson2015Read}. 

For an amplitude encoder, each component of $\mathbf{v}$ is then represented by the amplitude of the $n$-qubit ket vector $|\psi_{\text{in}}\rangle$ represented in computational basis. Without loss of generality, we assume that $m=2^n$ is a power of 2, otherwise we can attach $2^n-m$ zeros to the end of the vector $\mathbf{v}$ so that it can be transformed into a $n$ qubit pure state. The encoder can be realized by a circuit and the depth of the circuit is linear with the number of features \cite{mottonen2004quantum,knill1995approximation,plesch2011quantum}. 
Under certain conditions, a polynomial size of gate complexity over $m$ might be needed \cite{grover2002creating,soklakov2006efficient}. Such encoding procedure can be improved using a more complex approach like tensorial feature maps \cite{schuld2020circuit}. 

\subsection{III. The training process of quantum classifiers}

In classical machine learning, different loss functions are introduced when training the networks and estimating the performance. In numerical simulations, we employ a quantum version of cross-entropy as
\begin{equation}\label{eq:lossfunction}
\mathcal{L}(h(|\psi\rangle;\Theta),{\bf p})=-\sum_{i=1}^2 p_k\log q_k,
\end{equation}
where ${\bf q}=(q_1,q_2)$ is the diagonalized expression of output state $\text{diag}(\rho_{\text{out}})$ and ${\bf p}=(1,0)$ for $y=0$ and ${\bf p}=(0,1)$ for $y=1$. In the training procedure of a quantum classifier, a optimizer is used to adjust the parameter $\Theta$ to minimize the empirical loss function $\mathcal{L}_N(\theta)=\frac{1}{N}\sum_{i=1}^N\mathcal{L}(h(|\psi_i\rangle;\Theta),{\bf p}_i)$.
In recent years, a large family of gradient-based algorithms have been broadly used in training classical and quantum neural networks\cite{wilde2020stochastic,yamamoto2019natural,stokes2019quantum,kingma2014adam,sashank2018convergence}. In the numerical simulations in this research, we use Adam optimization algorithm \cite{kingma2014adam,sashank2018convergence}, which is a gradient-based learning algorithm with adaptive learning rate.

\begin{figure}
\hspace*{-0.24\textwidth}
\includegraphics[width=.24\textwidth]{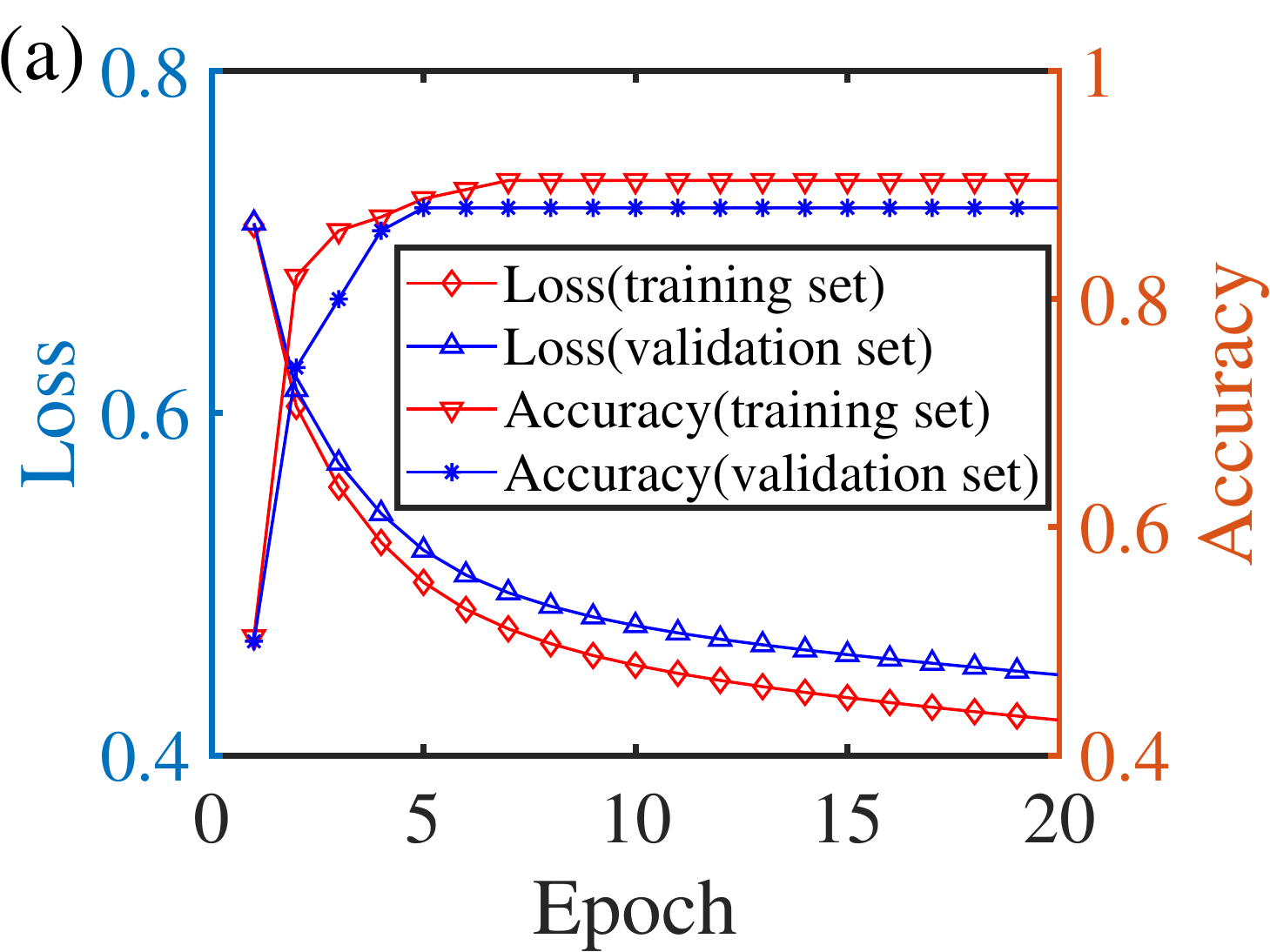}
\includegraphics[width=.24\textwidth]{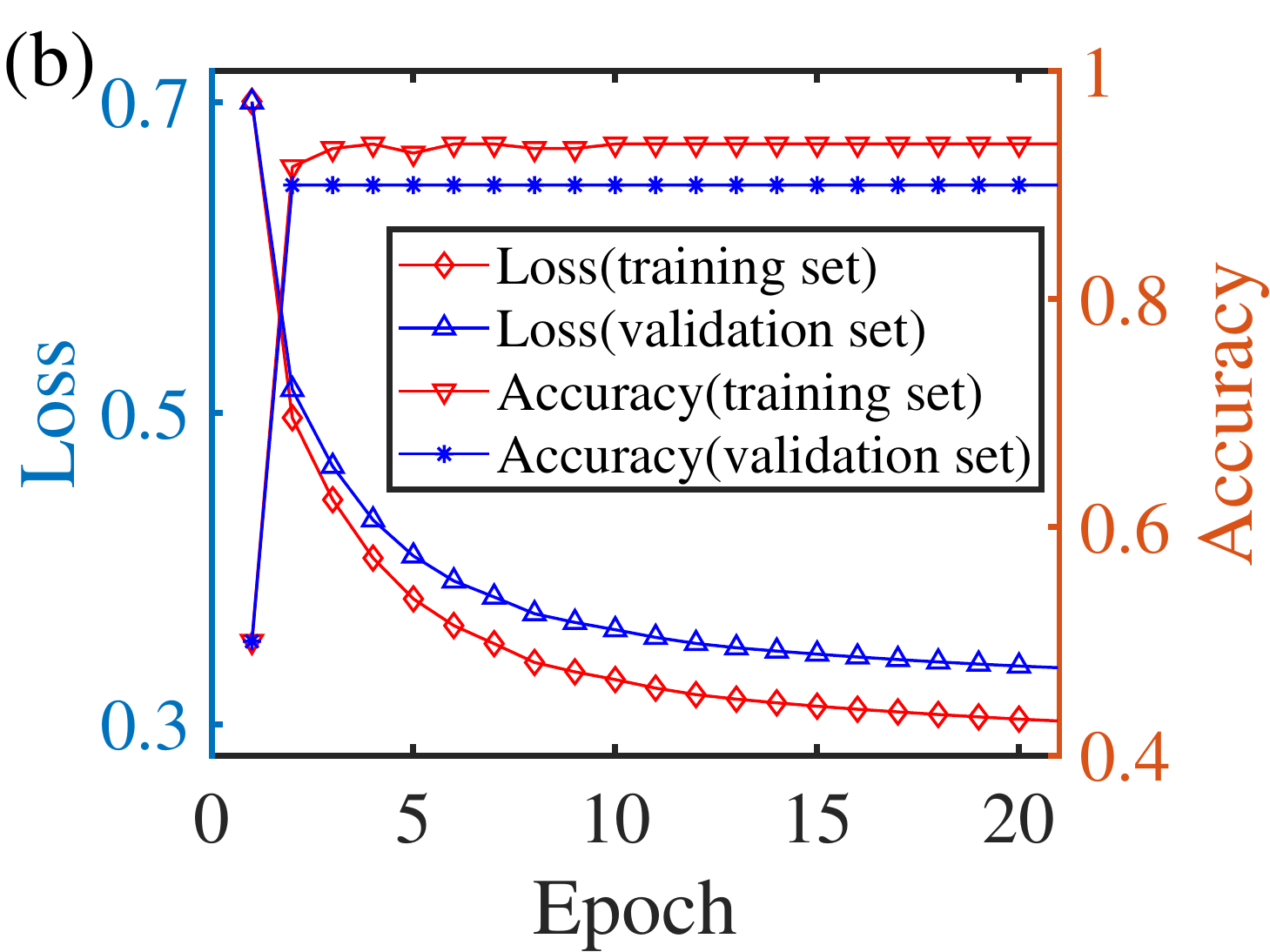}\\
\hspace*{-0.24\textwidth}
\includegraphics[width=.24\textwidth]{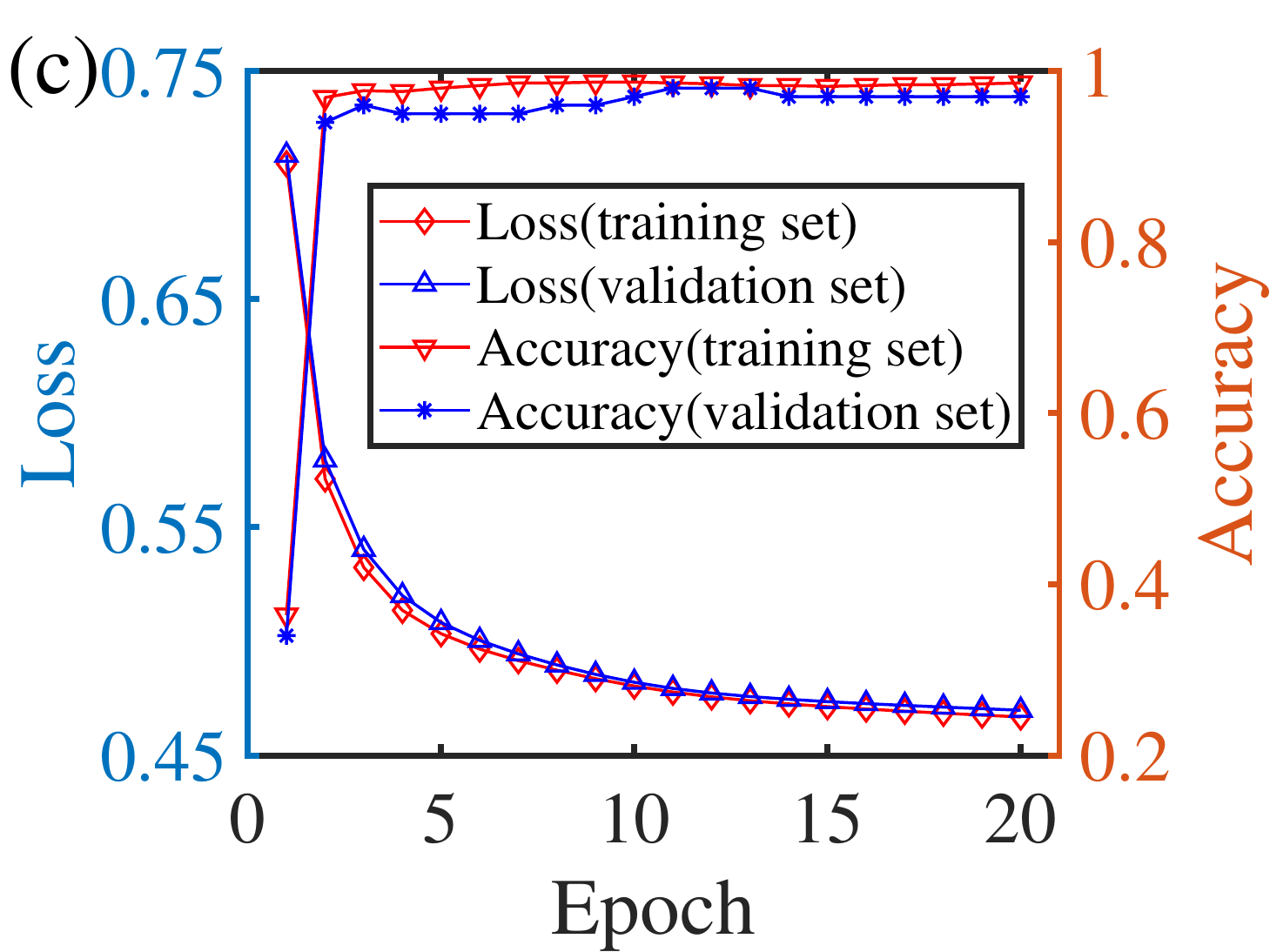}
\includegraphics[width=.24\textwidth]{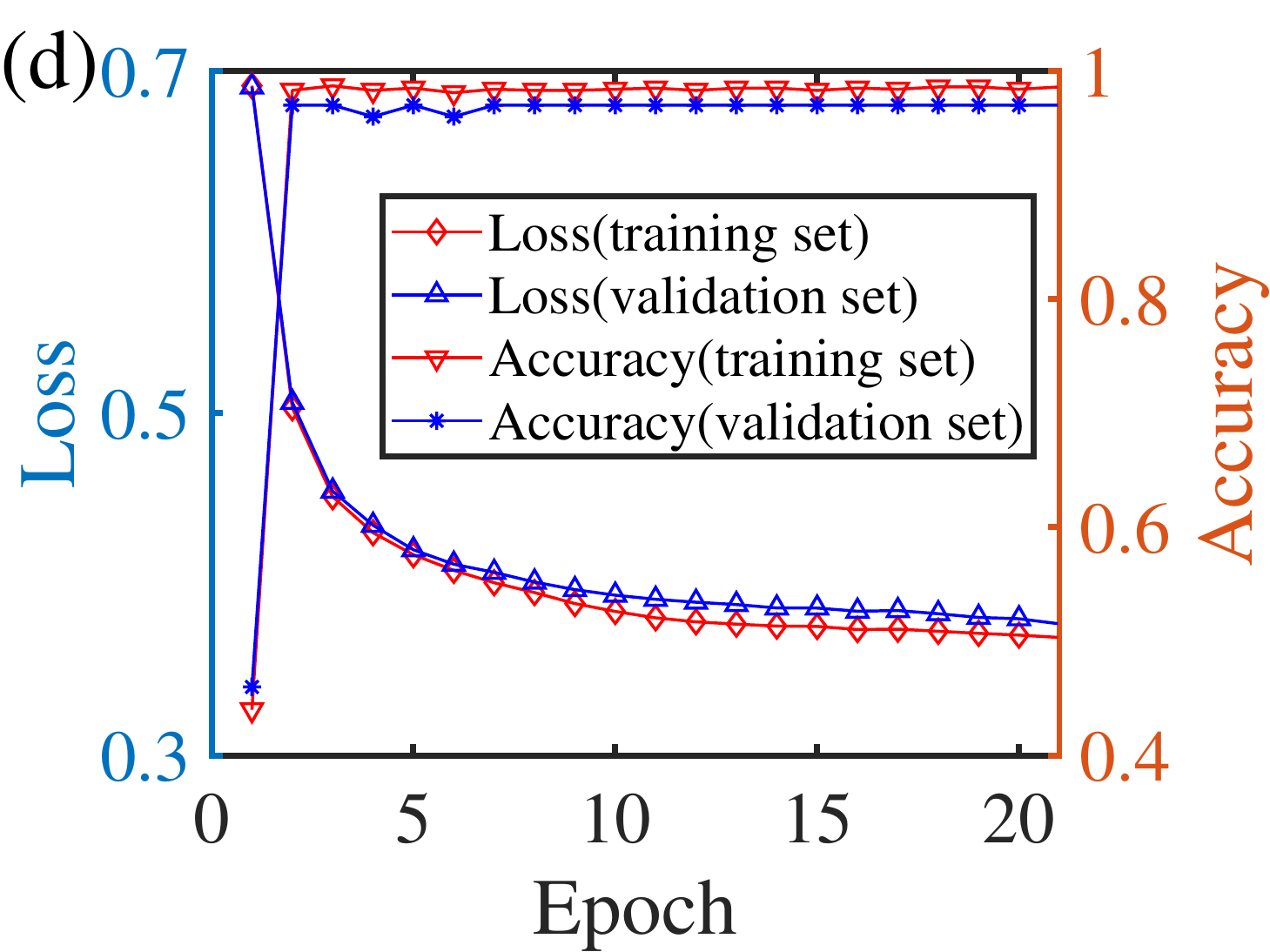}\\
\caption{The average loss and accuracy for quantum classifiers $2$ and $8$ during the training process, in classifying handwritten digit images and the ground states of the 1D transverse field Ising model.  
(a) The training procedure of the classifier $2$ (a QCNN classifier) for classifying the ground states.   Each epoch contains $30$ iterations. (b) The training procedure of the classifier $8$ with depth $p=10$ for classifying the ground states. Each epoch contains $5$ iterations.
(c) The training procedure of the classifier $2$  for classifying handwritten digit images. Each epoch contains $50$ iterations. (d)The training procedure of classifier $8$  for classifying handwritten digit images.  Each epoch represents $10$ iterations.}
\label{training}
\end{figure}

To find the minimization of the loss function using multiple-step gradient-based methods, we need to calculate the gradient of $\mathcal{L}_N(\Theta)$ over parameter $\Theta$. Each component of the gradient is represented as $\frac{\partial\mathcal{L}_N(\Theta)_\theta}{\partial\theta}=\lim_{\epsilon\rightarrow 0}\frac{1}{2\epsilon}[\mathcal{L}_N(\Theta)_{\theta+\epsilon}-\mathcal{L}_N(\Theta)_{\theta-\epsilon}]$ where $\theta$ is one of the parameters of $\Theta$. 
Owing to the special structures of the quantum classifiers,  we use the "parameter shift rule" \cite{liu2018differentiable,harrow2019low, lu2020quantum} in our numerical simulations to obtain the gradients required. 

In Fig. \ref{training}, we plot the average loss and accuracy of some of quantum classifiers in our classifier set during the training procedure. The numerical simulations including the training procedure and adversarial attack were done on a classical cluster using  Yao.jl\cite{Yao} and its extension packages in Julia language\cite{bezanson2017julia}. To run the simulation on GPU, i.e. to perfectly fit the mini-batch gradient descent algorithm, we use CuYao.jl\cite{CuYao}. This package is an efficient extension of Yao.jl on GPU calculation that can obtain a speedup. Flux.jl\cite{innes2018flux} and Zygote.jl\cite{zygote} packages are used to calculate the differentiation of the function. We note that the overfitting risk is low as the loss of the training data and validation data is close.\cite{srivastava2014dropout}.

In Table \ref{tab}, we list the number of parameters for each quantum classifier used in this paper, and their final accuracy  in classifying the ground states of the 1D transverse field Ising model. 

\begin{table}[H]
\centering
\begin{tabular}{|r|r|r|r|}
\hline
\hline
Classifier & Structure & Number of parameters & Accuracy\\
\hline
1 & QCNN & 44 & 0.917 \\
\hline
2 & QCNN & 92 & 0.950 \\
\hline
3 & Variational Circuit & 144 & 0.923 \\
\hline
4 & Variational Circuit & 171 & 0.930 \\
\hline
5 & Variational Circuit & 198 & 0.940 \\
\hline
6 & Variational Circuit & 225 & 0.930 \\
\hline
7 & Variational Circuit & 252 & 0.947 \\
\hline
8 & Variational Circuit & 279 & 0.955 \\
\hline
\hline
\end{tabular}
\caption{The number of parameters  and the final accuracy after the training process for each quantum classifier in classifying the ground states of the 1D Ising model. The accuracy is calculated over a training set that contains $300$ samples. 
}\label{tab}
\end{table}

\section{D. Adversarial algorithms}

In this section, we provide more details on the algorithms for obtaining adversarial examples and perturbations.

When proposing an adversarial attack on a quantum classifier that takes quantum data as input, we maximize the adversarial risk $\mu(\mathcal{E})$ mentioned in the main text. However, in practice $\mu(\mathcal{E})$ is typically inaccessible.  Hence, we consider maximizing the loss function instead. It is worthwhile to mention that a maximal loss function value does not always indicates a maximal risk. In the quantum scenario, we denote the adversarial perturbation attached to the quantum sample as an operator $U_\delta$ that acts on the input state. The maximization problem of adding perturbation can be described as:
\begin{equation}\label{perturbation}
U_\delta\equiv\mathop{\arg\max}_{U_\delta\in\Delta}\mathcal{L}(h(U_\delta|\psi\rangle;\Theta^*),{\bf p}),
\end{equation}
where $\Theta^*$ denotes the optimized parameters after the training process, $\Delta$ are the possible perturbations that can be added, $|\psi\rangle$ is the original input state and ${\bf p}$ is the correct label. In the case of studying universal adversarial examples, we have a test set $\mathcal{T}_M=\{(|\psi_0\rangle,y_0),...,(|\psi_M\rangle,y_M)\}$ and a set of quantum classifiers which learn hypothesis functions $h_1,h_2,...,h_k$. In order to obtain universal adversarial examples that can deceive all the quantum classifiers in the set, we solve the following optimization problem:   
\begin{equation}\label{perturbation:exp1}
U_{\delta}^j\equiv\mathop{\arg\max}_{U_\delta^j\in\Delta}\sum_{i=1}^{k}\mathcal{L}(h_i(U_\delta^j|\psi_j\rangle;\Theta^*),{\bf p}_j),
\end{equation}
where $U_\delta^j$ is the perturbation for the $j$-th sample. For the case of obtaining universal adversarial perturbations, we use an identical perturbation to implement the adversarial attack to all samples in the test set $\mathcal{T}_M$. In this case, we have one quantum classifier and its hypothesis function $h$. The maximization problem can be expressed in the similar form
\begin{equation}\label{perturbation:exp2}
U_{\delta}\equiv\mathop{\arg\max}_{U_\delta\in\Delta}\frac{1}{M}\sum_{i=1}^{M}\mathcal{L}(h(U_\delta|\psi_i\rangle;\Theta^*),{\bf p}_i).
\end{equation}

In general, the set $\Delta$ can be the set of unitary operators that are close to the identity matrix. We use automatic differentiation \cite{rall1996introduction} to improve precision when  applying the perturbation. In practice, we restrict the set $\Delta$ to be the product of local unitary operators near the identity matrix.

In the white-box attack scenario, the attacker has the full information of the classifiers, including their inner structures and loss functions. The attacker can then calculate the gradient of loss functions $\nabla\mathcal{L}(h(|\psi\rangle;\Theta^*),{\bf p})$. In this scenario, we use the quantum-adapted Basic Iterative Method (qBIM) introduced in Ref. \cite{lu2020quantum} to solve the above optimization problems in Eq. \eqref{perturbation:exp1} and \eqref{perturbation:exp2}. 

Compared with a white-box scenario, the adversary in a black-box setting does not have complete information of the quantum classifier. In classical adversarial learning, the development of black-box assumption has been divided into several categories. In non-adaptive black-box attack \cite{papernot2016transferability,papernot2017practical,tramer2016stealing}, the adversary knows nothing about the classifier's inner structure but can get access to the training data and analyze its distribution. In adaptive black-box scenario \cite{fredrikson2015model,papernot2017practical,rosenberg2017generic}, the attacker can use the classifier as an oracle without extra information provided. Another category is the strict black-box scenario \cite{hitaj2017deep}, where the data distribution is unknown but the adversary can collect the input-output pairs from the target classifier. 
In our simulations, we implement the non-adaptive black-box adversarial attack in which we try to use the knowledge of one quantum classifier to attack all quantum classifiers in the set that share the same training set and test set. The result shown in Fig. 2(c) in the main text indicates the effectiveness of such a black-box attack. 
\end{document}